\documentclass[aps,prb,floats,superscriptaddress,onecolumn,preprint]{revtex4-1}
\usepackage{amssymb}
\usepackage{amsmath}\usepackage{graphicx}
\usepackage{color}

\graphicspath{{../Figure.d/}}

\usepackage{color}

\begin{document}
\title{Revealing the role of organic cations in hybrid halide perovskite CH$_{3}$NH$_{3}$PbI$_{3}$}

\author{Carlo Motta}
\email[]{mottac@tcd.ie}
\affiliation{School of Physics, AMBER and CRANN Institute, Trinity College, Dublin 2, Ireland}
\affiliation{Qatar Environment and Energy Research Institute, P.O. Box  5825 Doha, Qatar}
\author{Fedwa El Mellouhi}
\email[]{felmellouhi@qf.org.qa}
\affiliation{Qatar Environment and Energy Research Institute, P.O. Box  5825 Doha, Qatar}
\author{Sabre Kais}
\affiliation{Qatar Environment and Energy Research Institute, P.O. Box  5825 Doha, Qatar}
\affiliation{Department of Chemistry and Physics, Purdue University, USA}
\author{Nouar Tabet}
\affiliation{Qatar Environment and Energy Research Institute, P.O. Box  5825 Doha, Qatar}
\author{Fahhad  Alharbi}
\affiliation{Qatar Environment and Energy Research Institute, P.O. Box  5825 Doha, Qatar}
\author{Stefano Sanvito}
\affiliation{School of Physics, AMBER and CRANN Institute, Trinity College, Dublin 2, Ireland}

\begin{abstract}
The hybrid halide perovskite CH$_{3}$NH$_{3}$PbI$_{3}$ has enabled solar cells to reach an efficiency of about 18\%, 
demonstrating a pace for improvements with no precedents in the solar energy arena. Despite such explosive progress, the 
microscopic origin behind the success of such material is still debated, with the role played by the organic cations in the 
light-harvesting process remaining unclear. Here van-der-Waals-corrected density functional theory calculations reveal that 
the orientation of the organic molecules plays a fundamental role in determining the material electronic properties. For instance, 
if CH$_{3}$NH$_{3}$ orients along a (011)-like direction, the PbI$_{6}$ octahedral cage will distort and the band gap 
will become indirect. Our results suggest that molecular rotations, with the consequent dynamical change of the band structure, 
might be at the origin of the slow carrier recombination and the superior conversion efficiency of CH$_{3}$NH$_{3}$PbI$_{3}$.

\end{abstract}

\date{\today}

\maketitle
\section*{\label{sec:intro} Introduction}
Very few materials have taken a research field by storm as the hybrid halide perovskites have done in the last two 
years in the solar cell community~\cite{Snaith:2013}. These are compounds with the standard AMX$_3$ perovskite 
structure, where X is the halide iodine, M is lead, while the remaining cation position, A, is taken by an
organic molecule, in this case methylammonium, CH$_{3}$NH$_{3}$. The success of such compounds is that 
high-efficiency solar cells can be fabricated cheaply from the liquid phase~\cite{Boix:2014,Loi:2013} and that the 
efficiency can be tuned by controlling their structural order and composition~\cite{Deschler:2014, Mosconi:2013}. 
Particularly intriguing is the fact that high efficiencies are achieved even for planar cells~\cite{Liu:2013}, indicating 
that charge separation can occur in the hybrid perovskite absorber and that efficient charge diffusion takes place for 
both electrons and holes~\cite{Xing:2013}.

Part of this behaviour can be related to the unusual situation concerning possible intrinsic defects. Recent 
electronic structure calculations~\cite{Yin:2014,Kim:2014} have shown that the dominant simple or complex 
defects in CH$_{3}$NH$_{3}$PbI$_{3}$ produce only shallow defect levels, while those able to act as deep 
traps have a high formation energy. Such observation suggests that, under ideal conditions, deep-level point 
defects cannot act as non-radiative recombination sites; a fact that can partially explain the relative long exciton 
lifetimes and diffusion lengths observed in these materials. At the same time the absence 
of deep levels may explain also the large open-circuit voltage measured. However, despite all these hints many 
fundamental questions remain open. In particular it is still not clear what is the role played by the organic molecules in the 
efficiency of these compounds. Theoretical investigations have now built a consensus that the 
methylammonium (MA) group does not have any significant contribution to the electronic structure around 
the band edges~\cite{Yin:2014,Brivio:2014,Umari:2014} and suggest that the only role of the molecules is that 
of donating an electron to the Pb-I framework, thus stabilizing the perovskite structure.

Interestingly, not even the geometrical arrangement of the MA group is known in detail. At low temperature 
(below 150~K) the crystal adopts an orthorhombic ({\it Pnma}) structure, in which the PbI$_{6}$ octahedra are 
strongly deformed assuming a rectangular basal plane. Such deformation restricts the rotational degrees of 
freedom of MA in the rhombus-shaped interstitial region, thus imposing a spatial ordering to CH$_{3}$NH$_{3}$. 
In this case the organic cation is pinned and can only rotate along the C-N axis. As the temperature is increased 
and the material progressively assumes a cubic structure ({\it Pm-3m}), by passing through a tetragonal one 
({\it I4/m}), the CH$_{3}$NH$_{3}$ molecules become free to rotate between the octahedral cages. Above room 
temperature such rotation is fast, to a point where both crystallographic analysis~\cite{Loi:2013, Baikie:2013} 
and NMR measurements~\cite{Knob:1990} have shown that the exact location of the MA groups cannot 
be determined.

Here we conduct  density functional theory (DFT) calculations for CH$_{3}$NH$_{3}$PbI$_{3}$ 
in the cubic phase, by taking into consideration the experimentally reported evidence of the fast rotation of 
CH$_{3}$NH$_{3}$. This means that we do not limit our analysis to CH$_{3}$NH$_{3}$ oriented along the 
(100) or (111) direction, for which the high O$_h$ symmetry is maintained, but we also explore cases where 
the symmetry is lowered. We find that such symmetry lowering has profound consequences on the electronic 
structure, namely that the bandgap changes from direct to indirect. Crucially such symmetry-lowering 
configurations represent local minima in the free energy surface of the crystal and they are stabilized
by van der Waals (vdW) interactions. These are the key ingredient not only for obtaining accurate lattice 
parameters\cite{Baikie:2013,Yun:2014,PhysRevB.90.045207,Egger:2014,Kronik:2014} (see Supplementary Figure~1) 
but also for the internal geometry. Our calculations then return a 
picture of the CH$_{3}$NH$_{3}$PbI$_{3}$ as a ``dynamical''  bandgap semiconductor, in which 
the exact position of the conduction band minimum depends on the particular spatial arrangement of the 
molecules. Importantly our results are robust against bandgap corrections and spin-orbit 
interaction~\cite{Umari:2014, Even:2013, Even:2014, Feng:2014, Giorgi:2014}, and deliver an absorption spectrum 
in good agreement with experimental data near the absorbtion edge.~\cite{Xing:2013,Lee_Teuscher_Miyasaka_Murakami_Snaith_2012}.

\section*{\label{sec:results} Results}
\subsection*{Geometrical Optimization}
Let us start the discussion by presenting the relaxed crystal geometries. Initially all the calculations have been 
performed at the level of the generalized gradient approximation (GGA)~\cite{Perd96} to the DFT exchange 
and correlation functional, including vdW interactions~\cite{PhysRevLett.102.073005} (see computational details section) 
and without explicitly considering spin-orbit coupling (SOC). Additional calculations beyond the GGA and including SOC 
will be discussed later in relation to the electronic structure.

\begin{figure}
\includegraphics[width=0.7\columnwidth]{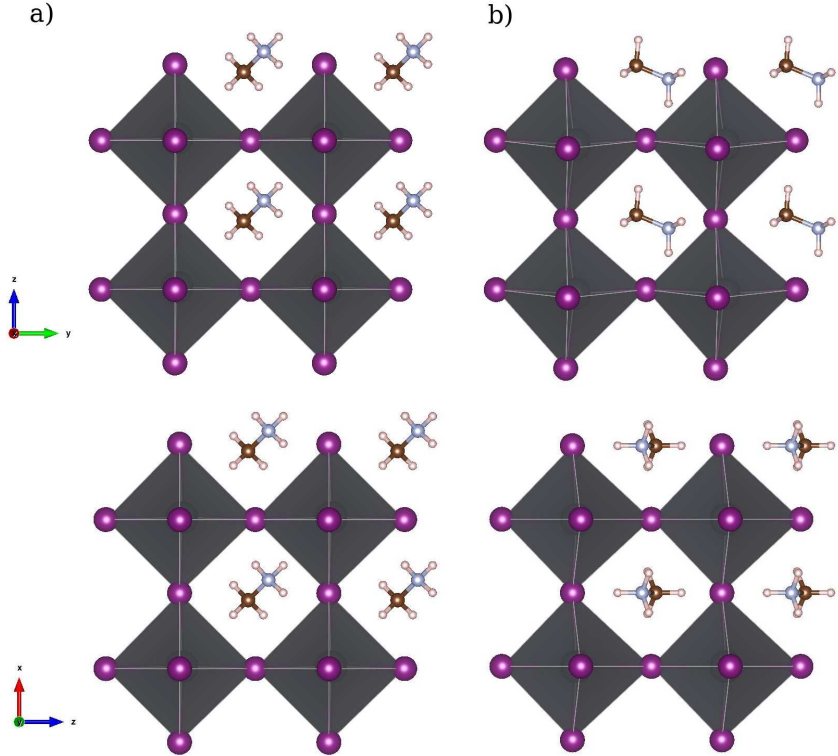}
\caption{\label{fig:1} Relaxed structures of the cubic phase of CH$_{3}$NH$_{3}$PbI$_{3}$ for two different
orientations of the cations, namely along (a) (111) and (b) (011) (b). The views along the $xy$ and $xz$ 
planes are shown in the upper and lower panels, respectively.}
\end{figure}

In Fig.~\ref{fig:1} the relaxed structures for the cubic phase of CH$_{3}$NH$_{3}$PbI$_{3}$ are presented along 
the $xy$ and $xz$ planes for two different orientations of the CH$_{3}$NH$_{3}$ cation. The relaxation process 
is extremely sensitive to the initial conditions. When we start the geometrical optimization with the molecular cation
oriented along the (111) direction, the structure relaxes maintaining the same orientation [panel (a)]. In contrast, 
by starting from a (001) oriented molecule the relaxation may end up with CH$_{3}$NH$_{3}$ along (011) 
[panel (b)] or an equivalent directions. However, it is important to point out that the cation energy landscape is very 
shallow. Indeed, both configurations represent local energy minima, with an energy difference of only 
about $20$~meV in favor of (011). In both cases, the lattice parameter is calculated in excellent agreement 
with the experimental value of 6.329\AA~\cite{Poglitsch1987} and an error of less than 
0.8\% (see Supplementary Table~1 and Supplementary Figure~1). 
If one does not include vdW forces in the relaxation, the discrepancy with the 
nominal value will become larger than 2\%, i.e. the crystal structure is not properly described. Furthermore, the molecular 
bending angle increases from $16^{\circ}$ to $25^{\circ}$ upon inclusion of dispersion forces. In order to make a 
well-grounded comparison with previous calculations~\cite{Umari:2014, Brivio:2013} reporting no effect of 
methylammonium on the structure, we started the relaxation from the experimentally published atomic 
coordinates with CH$_{3}$NH$_{3}$ along (100)~\cite{Stoumpos:2013}, whose band structure is presented in
Supplementary Figure~2.
%
%

By inspecting the relaxed structures a prominent difference between the (111) and the other two geometries
appears. In fact, in the case of (011) the rotation of the cation  induces a deformation and a symmetry reduction 
of the inorganic PbI$_6$ octahedra. The result of such PbI$_6$ octahedra distortion is that the Pb-I bonds 
do not lie parallel to the crystal directions, but instead they slightly deviate by approximately $6^{\circ}$. Unlike 
the (111) case, the unit cell becomes pseudocubic with the lattice parameters along the $y$ and $z$ directions 
being 1\% larger than that along $x$.  Hence, it is clear that the dispersive forces are critical for the internal 
geometry optimization and consequently, as we will see next, for the electronic behavior of hybrid perovskites.

\subsection*{Electronic Properties}
\begin{figure}
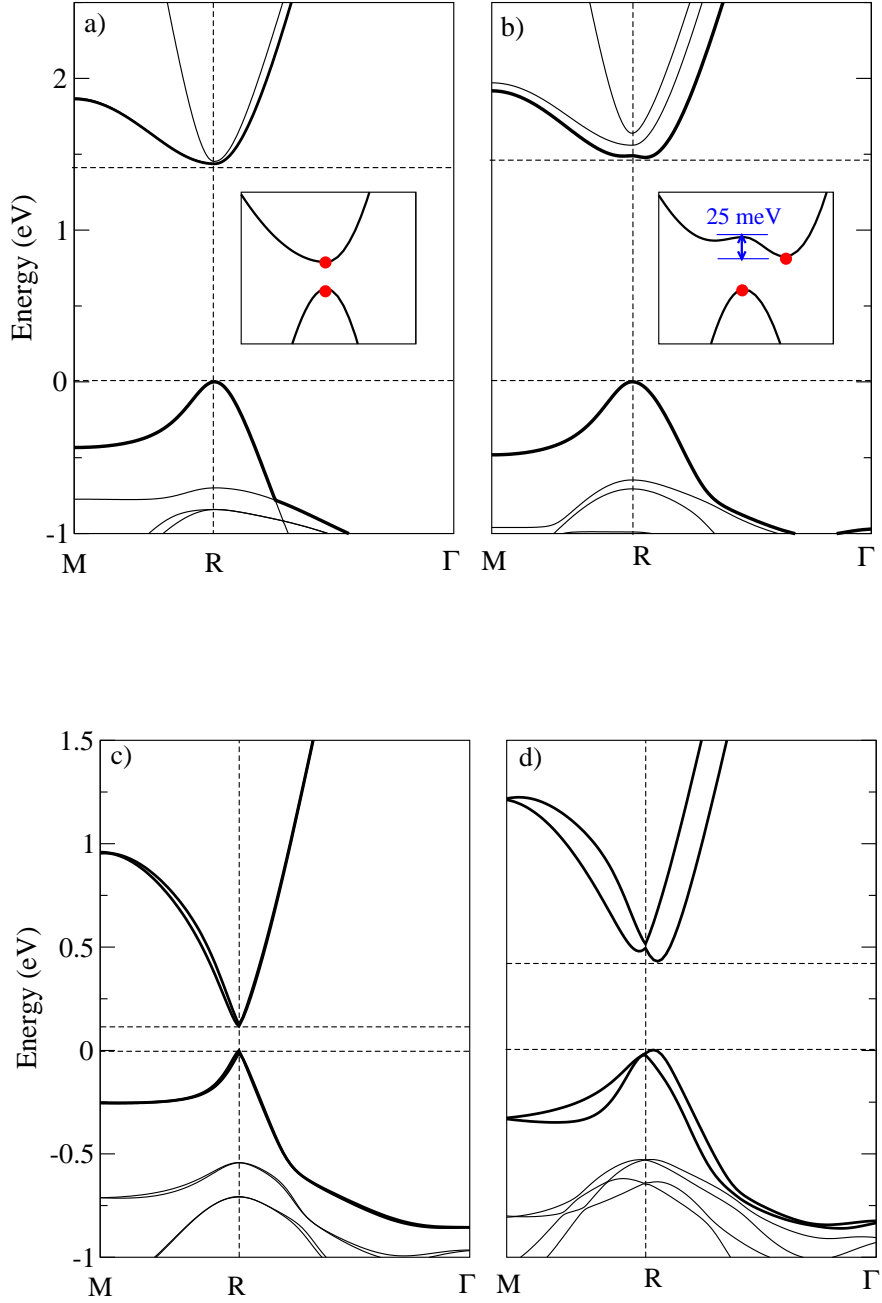

\includegraphics[width=0.7\columnwidth]{fig2}
 \vspace{2cm} \\
\includegraphics[width=0.7\columnwidth]{fig3}
\caption{\label{fig:2} Band structure of the fully relaxed CH$_{3}$NH$_{3}$PbI$_{3}$ crystal for 
molecule orientation along (111), (a) and (011) (b). The insets show a magnification
of the bands (which have been shifted in energy for convenience) around the bandgap and 
highlight the changes in the VBM and CBM caused by the rotation of 
CH$_{3}$NH$_{3}$. Note that for the (011) orientation the bandgap becomes indirect. 
The relativistic SOC bandstructures are shown in panels (c) and (d) for (111) and (011).}
\end{figure}

The fine differences in the crystal geometries have a qualitatively strong impact on the electronic structure.
In Fig.~\ref{fig:2} we show the electronic bandstructure along some high symmetry points of the Brillouin zone
calculated with and without including SOC.
For both orientations the energy gap computed without SOC lies in the 1.5-1.7~eV range, close to the experimental 
value of about $1.55$~eV~\cite{Kojima:2009,Lee_Teuscher_Miyasaka_Murakami_Snaith_2012}. However, as already 
shown~\cite{Mosconi:2013}, this agreement is due to a fortuitous cancellation of errors, namely the 
GGA underestimates the bandgap, but this is counterbalanced by the lack of spin-orbit interaction. It is 
well known that the bottom of the conduction band mainly originates from the $p$ orbitals of Pb, while 
the top of the valence band is derived from the $p$ orbitals of I. The highest occupied molecular orbital 
(HOMO) of methylammonium is found deep below the valence band, approximately 5~eV below the 
valence band maximum (VBM). Thus, one may argue that CH$_{3}$NH$_{3}$ does not play any role 
in the optical and electronic response of such materials, but that rather it does only contribute to their 
structural cohesion. However, a careful analysis of the density of states (DOS) projected on the various atoms
[see Fig.~\ref{fig:3}(a) and Supplementary Figure~3] reveals that there is a small contribution $\sim0.5$~eV 
below the VBM attributable to the organic molecules. This indicates that indeed there is interaction between 
CH$_{3}$NH$_{3}$ and the inorganic PbI$_6$ octahedra, in the form of hydrogen bonds between N and I. 

A closer inspection to the bandstructures reveals that the orientation of CH$_{3}$NH$_{3}$ has a profound
impact on the nature of the bandgap. In fact, while for the (111) case the gap is direct at the R point, the 
cationic rotation induces a transition to an indirect gap, and when MA is along (011) the conduction band 
mininum (CBM) shifts along the $\mathrm{R}\rightarrow\Gamma$ line. Accordingly, the vertical energy difference 
between the CBM and the conduction band energy at R is 25~meV, as shown in the inset of Fig.~\ref{fig:2}(b). 
The effect persists when SOC is turned on, as shown in Fig.~\ref{fig:2}(c) and Fig.~\ref{fig:2}(d). In this case, 
a slight shift of the VBM towards $\Gamma$ occurs as well, reducing the CB vertical energy difference 
approximately to half the non-SOC value.
We stress here that the (011) orientation is not a special case: the indirect bandgap does appear for all
the equivalent orientations, namely (0$\pm$11), (01$\pm$1), ($\pm$101), ($\pm$110), (10$\pm$1), (01$\pm$1).
As an example, in Supplementary Figure~4 we show the (0-11) case, for which the CBM shifts in the direction opposite to that 
shown in Fig.~\ref{fig:2}, namely $\mathrm{R}\rightarrow\mathrm{M}$.

Such gap transformation is the most remarkable result of the interaction between the organic molecules
and the inorganic framework, which is driven by the dispersive forces. As a result, CH$_{3}$NH$_{3}$ exerts 
strain on the PbI$_6$ octahedra and changes the nature of the bandgap. This is unusual in the perovskites world
and it is a unique peculiarity of such hybrid systems. Typical inorganic perovskites such as SrTiO$_3$ respond 
to very small uniaxial strains, as those induced by the antiferrodistortive phase transition, by splitting their 
3-fold degenerate CBM~\cite{El-Mellouhi:2013}. To the best of our knowledge very few semiconductors see 
their bandgap changing from direct to indirect under the effect of strain. An example is given by layered transition
metals chalcogenides, like for instance MoSe$_2$ single-layer, where the gap change is attributed to phonon 
softening~\cite{Horzum:2013}. Intriguingly also for CH$_{3}$NH$_{3}$PbI$_{3}$ we observe some softening of our calculated  
low-energy optical phonon modes, when the methylammonium molecule rotates along the (011) directions. In
particular the calculated energy red shift is of the order of $10$~cm$^{-1}$ (see Supplementary Figure~7).

We have carefully verified that the indirect bandgap is not an artifact of our computation method. Firstly, we 
analyzed the relaxation in a 2$\times$2$\times$2 (96 atoms) supercell to establish whether further low-symmetry
configurations are possible. The supercell converged geometries are very similar to the unit cell geometries repeated 
in space due to periodic boundary conditions. The calculations reveal no particular changes, although the size of the 
supercell would allow the PbI$_6$ octahedra to rotate or tilt. Consequently, the features of the 
bandgaps are preserved both at the qualitative as well as the quantitative level. 

Secondly, we have tested different vdW functionals by comparing the geometries obtained with the Tkatchenko-Scheffler (TS)
scheme~\cite{PhysRevLett.102.073005} to those calculated with the empirical Grimme's method~\cite{JCCJCC20495} 
and with a recently developed sophisticated many-body dispersion (MBD) correction~\cite{PhysRevLett.108.236402}. As 
shown in Supplementary Figure~1, the unit cell volume compares well with the experimental value for all the vdW-corrected functionals, 
although that calculated with the TS scheme is closer. Thus, in general, the TS and Grimme's pairwise dispersion methods 
provide a sufficiently good description of our system as found in a recent work on molecular crystals~\cite{Kronik:2014}.
In any case the indirect bandgap is found for all the calculated geometries, including those obtained with the MBD correction.
In contrast, when dispersion interactions are not included the gap remains direct since the structure does not present 
enough distortion. Then, we have performed calculations using  the HSE06~\cite{Krukau:2006} screened hybrid functional 
by starting from the optimal relaxed structures. Also in this case the bandgap remains indirect 
(see Supplementary Figure~3), although 
the overall energy distance between the CBM and the VBM increases, as expected. 

Finally, we have performed additional calculations to establish whether or not the observed effect is the result of the interaction 
of the MA dipole with its periodic images. We have then verified that the position of the VBM and the CBM does 
not shift (i) upon flipping the molecule's dipole moment, and (ii) upon applying a rigid rotation of the molecule from the 111 
to the 110 orientation (Supplementary Figure~5). This confirms that it is the distortion of the inorganic PbI$_6$ octahedra to 
govern the shape of the CBM and VBM and not the dipole moment of the organic molecule.

As a last element of our analysis we have investigated the relative stability of the different molecular orientations. 
In particular, we have performed nudged elastic band calculations and evaluated the barrier height for a MA rotation 
from (011) to (111), taking into account possible changes in the cell geometry and the lattice vectors. We have
found a barrier of 20~meV (1.929~kJ/mol), which is less than $k_\mathrm{B}T$, and it is similar to values found 
recently in literature~\cite{Frost:2014}. This confirms the concept that above room temperature the molecules rotate 
at all time. Given the strong dependence of the position in $k$-space of the CBM and the VBM on the cell geometry, 
the fast molecular rotation implies that the nature and shape of the bandgap are rapidly-varying
functions of time. Recently Mosconi et al.~\cite{Mosconi:2014} published Car-Parinello molecular dynamics 
simulations for a 2$\times$2$\times$2 CH$_{3}$NH$_{3}$PBI$_{3}$ supercell at 319K, showing fluctuations of 
$\pm$0.1-0.2~eV in the position of the VBM, and thus giving evidence that the size of the bandgap 
changes as function of the MA rotation. Here we do a step forward and demonstrate that for some molecular orientations 
the bandgap turns from direct to indirect. 

\subsection*{Optical absorption}
\begin{figure}
\includegraphics[width=0.7\columnwidth]{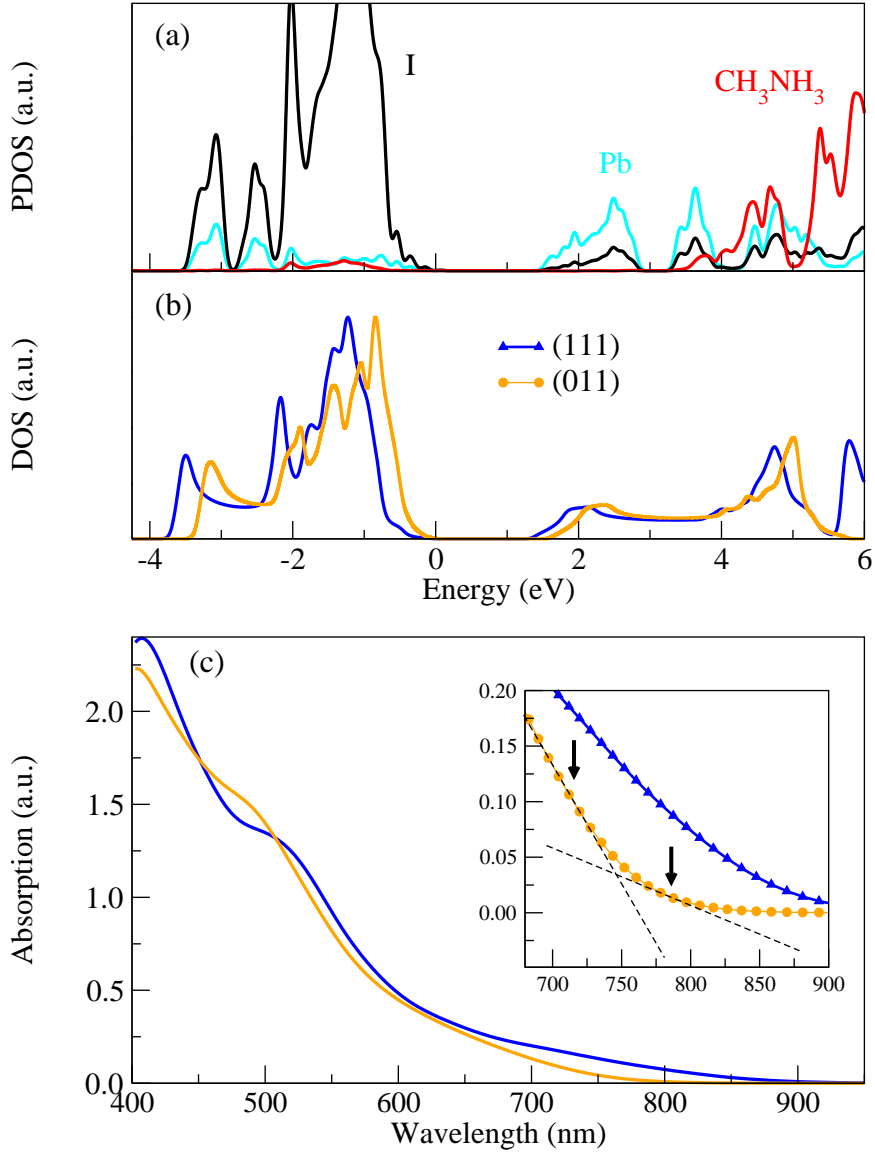}
\caption{\label{fig:3} Electronic and optical properties of the cubic phase CH$_{3}$NH$_{3}$PBI$_{3}$. 
In panel (a) we present the projected density of states for the case of the (111) oriented molecule on the 
Pb (cyan), I (black) species and the CH$_{3}$NH$_{3}$ (red). In panel (b) we display the density of states 
as a function of the chemical potential for the molecule oriented along the (111) and (011) directions.
In panel (c) the calculated absorption spectra are plotted for the different orientation of CH$_{3}$NH$_{3}$ 
clearly showing the impact the organic cation rotation. The inset in (c) is a magnification of the same quantity 
around the low-energy region.}
\end{figure}

In order to seek further evidence of the presence of the indirect bandgap, we have calculated the 
perovskite's optical absorption and compared it with available experimental data. The calculations
have been performed over the electronic structure obtained with the GGA including vdW interaction
in the independent particle approximation and without SOC. Although the fortuitous
cancellation of errors returns a bandgap close to experiments, we concentrate here on the shape
of the spectrum around the absorption edge. The calculations are performed for the two 
orientations of the CH$_{3}$NH$_{3}$ molecule, and the spectra are presented in Fig.~\ref{fig:3}(c).

When methylammonium is aligned along the (111) direction the bandgap is direct 
and measures 1.423~eV~=~871~nm. The absorption profile then increases monotonically from that 
value, indicating the possibility of direct transitions across the band edges. In contrast, when 
CH$_{3}$NH$_{3}$ is placed along (011) the bandgap is indirect, and in general 
slightly larger than that corresponding to the (111) orientation. In particular, for (011) we find a direct gap of 
1.611~eV~=~769~nm and an indirect one of 1.629~eV~=~761~nm. Here, the shape of the absorption spectrum
is quite different from the (111) case. We note a ``two-step'' absorption profile [see the inset of Fig.~\ref{fig:3}(c)], 
with a small absorption amplitude developing at the band edge followed by a significantly more pronounced 
increase at around 750~nm. This reflects the indirect nature of the bandgap and the fact that indirect transitions 
are not allowed, unless assisted by phonons. Certainly, the level of theory used here is not complete and one
should use many-body methods for deriving a more qualitative picture~\cite{Even:2014b}, however the inclusion 
of excitonic effects is not necessary to make our main point about the shape of the spectrum around the absorption 
edge. In fact the differences between the direct and indirect bandgap are very evident from the inset of Fig.~\ref{fig:3}(c), 
and should be compared with the typical experimental absorption spectrum measured at 300~K given 
in Supplementary Figure~6. 
This latter features a small amplitude around the bandgap followed by a rapid increase after about 50~nm, 
which seems to be a robust feature in the experimentally measured spectra, regardless of the morphology of the 
material~\cite{Xing:2013,Burschka:2013,Lee_Teuscher_Miyasaka_Murakami_Snaith_2012} and resistent 
to film quality degradation~\cite{C3TA13606J}.

Whether or not such near-edge feature can be attributed to the indirect bandgap is however difficult to establish
with certainty. In fact, at 300~K the MAs are free to rotate and sample the many quasi-equivalent (011) and (111) 
orientations. Thus, it is expected that no sharp indirect bandgap can be observed, but rather a smooth absorption 
spectra starting around the average bandgap value. Interestingly, we point out that our finding could help explaining 
the recently observed line broadening in photoluminescence experiments~\cite{doi:10.1021.jz500434p}, attributed to 
coupling with phonons. In fact, low energy phonons are associated with cation rotations, which in turn would shift the 
CBM to different positions. As a result, the range of optical gap would change as function of temperature thus broadening 
the emission spectrum. Measurement of the optical absorption and thermal characteristics using very high sensitivity 
techniques such as photothermal deflection spectroscopy might reveal the existence of the indirect bandgap by fine 
lowering of temperature below 300~K thus freezing the MA (011)-like orientations. 

\section*{\label{sec:discussion}Discussion}
Absorbing light efficiently and transporting the photocurrent with some gained voltage are the most
important aspects in the design of a solar cell. The two processes can be characterised, respectively, by 
an empirical absorption length, $L_\alpha$, inversely proportional to the absorption~\cite{fahad_a,fahad_b}, 
and by the carrier diffusion length, $L_{\rm diff}$. This is a measure of the mean distance travelled
by carriers before recombining~\cite{fahad_c,fahad_d,fahad_e}, it is directly proportional to 
the carrier lifetime, $\tau$, and it depends on several factors (mobility, carrier concentration, etc.).
In solar cells it is essential to fulfil the condition $r=L_{\rm diff} / L_\alpha > 1$ to ensure that the 
photogenerated carriers can be extracted and collected. For practical reasons $r\gg1$ is required
for high power conversion efficiency.~\cite{fahad_a,fahad_b,fahad_c,fahad_f}

In general, direct bandgap semiconductors, such as GaAs, have strong absorption and thus short $L_\alpha$, 
but this comes with fast carrier recombination and hence relatively small $L_{\rm diff}$. The situation is opposite 
when the bandgap in indirect (e.g. in Si). An ideal balance between  $L_\alpha$ and $L_{\rm diff}$ is present in 
some indirect bandgap semiconductors such as Cu$_3$N~\cite{fahad_b,fahad_c}, 
where the direct transition starts a few $k_B T$ above the indirect gap, yielding a short $L_\alpha$. The small difference 
between the indirect gap and the lowest direct transition ensures that the absorption is not greatly affected by
the indirect gap till the direct transitions start to dominate the absorption spectrum. In contrast, the lifetime of the
photogenerated carriers, which relax to the band edges in the indirect gap semiconductors, is longer than that 
of the direct ones by orders of magnitudes~\cite{fahad_d} ensuring a large $L_{\rm diff}$. This class of indirect 
semiconductors promises remarkable improvement in efficiency via the large 
$L_{\rm diff} / L_\alpha$ ratio~\cite{fahad_b,fahad_c}.

From this short discussion and our results at hand we can speculate on the possible origin of the high efficiency 
in hybrid perovskites solar cells. Our calculations establish that, under certain conditions of molecule alignment, 
the bandgap of the high-temperature phase of CH$_{3}$NH$_{3}$PBI$_{3}$ can be indirect and not always direct 
as previously believed. This feature originates from the strain exerted on the PbI$_6$ octahedra by the organic 
molecules, when these points mostly toward one of the unit cell faces rather than along the diagonal, which constitute 
shallow minima in the total energy profile. The indirect bandgap is tens of meV smaller than the direct one and the 
position of the CBM with respect to the VBM depends on the specific orientation of the molecules. This analysis delivers 
a picture of CH$_{3}$NH$_{3}$PBI$_{3}$ as a ``dynamical'' bandgap semiconductor, where the term ``dynamical'' refers 
to the fact that the position in $k$-space of the CBM depends on the molecular alignment, which in turn fluctuates at room 
temperature. The calculated absorption spectrum near the band edge, characterized by a shallow amplitude at about 
800~nm followed by a rapid enhancement within 50~nm, is very similar for (011) and equivalent orientations of the 
molecules, indicating that it is a feature robust to the molecular rotation and disorder.

The presence of an indirect bandgap, with only a small displacement in $k$-space with respect to the direct point, suggests 
that the rapid relaxation of the photo-excited electrons to the conduction band minimum is still possible. This process is 
incoherent, mediated by acoustic phonons, and spontaneously breaks the excitons in free carriers. As a second effect of 
the collapse of the electrons to the indirect conduction band minimum is that the lifetime of the minority carriers is enhanced, 
so that radiative recombination is partially suppressed. Such suppression is not complete since experiments reveal luminescence, 
however it contributes to prolong the carrier lifetime. In fact, it is interesting to note that experiments report that the 
luminescence gets suppressed as the material crystallizes~\cite{Burschka:2013} and that in any case it is sensitive to the 
sample morphology and quality~\cite{Choi:2013}. Considering that the crystallization process leads to the geometries described 
here, such suppression of the luminescence upon ordering supports our argument. To complete the picture it is also worth 
reminding that the absence of strong scattering centers inhibits incoherent recombination, so that  $L_{\rm diff}$ 
remains long~\cite{Stranks:2013}.

\section*{\label{sec:conclusions}Conclusions}
Despite the tremendous advances in studying and understanding hybrid perovskites from both the theoretical and the 
experimental point of views, yet the reason behind the extended carrier lifetime in these materials remains a mystery. 
Our work suggests that extended carrier lifetimes in CH$_{3}$NH$_{3}$PBI$_{3}$ might be a consequence of the  
``dynamical''  position  of the conduction band minimum forming a basin  around a high symmetry point in the $k$-space. 
This proposed dynamical bandgap, induced by  the molecular motion, prohibits carrier recombination, favors
exciton separation thus  increasing the conversion efficiency of the photvoltaic absorber  CH$_{3}$NH$_{3}$PBI$_{3}$. 
Crucially, all such phenomenology appears to be initiated by the geometry assumed by the molecules following full 
geometry relaxation. A set of differently oriented CH$_{3}$NH$_{3}$ cation geometries, relaxed without any symmetry 
constrains using DFT including van der Waals forces, lead to geometries where the PbI$_6$ octahedra 
distort. The band-structure corresponding to such geometries present an indirect bandgap, which persists upon  bandgap 
corrections via screened hybrid functionals and spin-orbit coupling. Finally, we propose that a possible way for
engineering the efficiency of hybrid perovskites is to induce distortion to the PbI$_6$ octahedra either by using strain or 
organic molecules strongly interacting with the inorganic cages.

\section*{\label{sec:compdet}Computational details}

\noindent
\textbf{DFT calculations} \\
The electronic properties of the hybrid perovskites have been calculated with the all-electron {\sc fhi-aims} code at the level
of GGA in the Perdew-Burke-Ernzerhof (PBE)~\cite{Perd96} parameterization. Long-range van der Waals interactions 
have been taken into account via the Tkatchenko and Scheffler (TS) scheme~\cite{PhysRevLett.102.073005}, which is
also constructed over a GGA and a pairwise dispersive potential. The reciprocal space integration was performed over 
an 8$\times$8$\times$8 Monkhorst-Pack grid~\cite{PhysRevB.13.5188}. A pre-constructed high-accuracy all-electron 
basis set of numerical atomic orbitals was employed, as provided by the {\sc fhi-aims} ``tight'' default option 
(more details are given in Supplementary Information). 
Unconstrained structural optimization was performed with the Broyden-Fletcher-Goldfarb-Shanno algorithm~\cite{numrec97} with 
with a tolerance of $10^{-3}$~eV/\AA\ . 

Additional calculations, in particular concerning the use of the hybrid functional HSE06~\cite{Krukau:2006}, were
performed using the {\sc Gaussian} suite\cite{g09}, with the periodic boundary condition~\cite{Kudin:2000} code 
version. The Def2-\cite{Weigend:2005} series of {\sc Gaussian} basis sets were optimized following our own procedure, 
described in Ref.~[\onlinecite{El-Mellouhi:2011}], for the atomic species of interest in this work. Most numerical 
settings in {\sc Gaussian} were left at the default values, \textit {e.g.}, uncontrained geometry optimization settings, integral 
cut-offs, and self-consistent convergence thresholds. The standard rms force threshold in GAUSSIAN for geometry
optimizations is 450$\times$10$^{-6}$  hartrees/bohr. The default $k$-point mesh of 8$\times$8$\times$8 was 
used for the 12 atoms unit cell of CH$_{3}$NH$_{3}$PbI$_{3}$. 

Finally calculations involving the spin-orbit interaction have been carried out  using the Vienna ab initio simulation package (VASP). 
A plane wave basis set energy cutoff of 520~eV was considered in the calculations and Brillouin zone integration was performed 
using 8$\times$8$\times$8 k-point mesh centered at the Gamma point. Dispersion correction to the PBE was done  within the method 
of Tkatchenko and Scheffler (DFT-TS) ---recently implemented in the 5.3.5 version of VASP ---  
used to account for Van der Waals interactions. The projector augmented wave (PAW) pseudopotentials were used  
for all elements  $5d^{10}6s^{2}6p^{2}$ valence electron potential was used for the Pb atom.


The linear optical absorption has been obtained with {\sc fhi-aims} by computing the momentum matrix elements
in a 5~eV energy window around the Fermi level and then by using the Fermi golden rule approximation.
A dense k-point sampling of 15$\times$15$\times$15 has been employed to ensure converged spectra.\\

\noindent
\textbf{Phonon calculations}\\
Phonons spectra were  computed with the frozen-phonon approach for a fully relaxed 2$\times$2$\times$2 supercells 
using  the Phonopy~\cite{phonopy} code in combination with {\sc fhi-aims}  A displacement of 0.001 \AA \ was applied to 
each atom in the three space directions to compute the dynamical matrix.The phonon density of states used the $\Gamma$ 
point sampling while the full phonon band structure was generated using  the $\Gamma$, $M$, $X$ and $R$ high symmetry 
points in the Brillouin zone. 
 

\section*{Acknowledgements}
This work is sponsored by the European Research Council, {\sc Quest} project, (CM and SS) and by
the Qatar Environment and Energy Research Institute (FE,  FHAH, NT and SK). Computational resources 
have been provided by the supercomputer facilities at the Trinity Center for High Performance 
Computing, at ICHEC (project tcphy038b) and the research computing at Texas A\&M University at Qatar. 
We aknowledge fruitful discussions with Simon Gelinas, Marcelo Carignano and Mohamed Lamine Madjet and 
thank  Felix Deschler for sending his absorbtion spectra. 

\section*{Author contributions}
C.M. and F.E.M. contributed equally to the manuscript. 
C.M. and F.E.M. performed the computational research, C.M., F.E.M., S.K., N.T., F.A. and S.S. discussed the results.
C.M., F.E.M. and S.S. contributed to the preparation of the manuscript.
\section*{Additional information}

\textbf{Competing financial interests}: The authors declare no competing financial interests.


\begin{thebibliography}{10}
\expandafter\ifx\csname url\endcsname\relax
  \def\url#1{\texttt{#1}}\fi
\expandafter\ifx\csname urlprefix\endcsname\relax\def\urlprefix{URL }\fi
\providecommand{\bibinfo}[2]{#2}
\providecommand{\eprint}[2][]{\url{#2}}

\bibitem{Snaith:2013}
\bibinfo{author}{Snaith, H.~J.}
\newblock \bibinfo{title}{Perovskites: The emergence of a new era for low-cost,
  high-efficiency solar cells}.
\newblock \emph{\bibinfo{journal}{J. Phys. Chem. Lett.}}
  \textbf{\bibinfo{volume}{4}}, \bibinfo{pages}{3623--3630}
  (\bibinfo{year}{2013}).

\bibitem{Boix:2014}
\bibinfo{author}{Boix, P.~P.}, \bibinfo{author}{Nonomura, K.},
  \bibinfo{author}{Mathews, N.} \& \bibinfo{author}{Mhaisalkar, S.~G.}
\newblock \bibinfo{title}{Current progress and future perspectives for
  organic/inorganic perovskite solar cells}.
\newblock \emph{\bibinfo{journal}{Mater. Today}} \textbf{\bibinfo{volume}{17}},
  \bibinfo{pages}{16--23} (\bibinfo{year}{2014}).

\bibitem{Loi:2013}
\bibinfo{author}{Loi, M.~A.} \& \bibinfo{author}{Hummelen, J.~C.}
\newblock \bibinfo{title}{Hybrid solar cells: Perovskites under the sun}.
\newblock \emph{\bibinfo{journal}{Nat. Mater.}} \textbf{\bibinfo{volume}{12}},
  \bibinfo{pages}{1087--1089} (\bibinfo{year}{2013}).

\bibitem{Deschler:2014}
\bibinfo{author}{Deschler, F.} \emph{et~al.}
\newblock \bibinfo{title}{High photoluminescence efficiency and optically
  pumped lasing in solution-processed mixed halide perovskite semiconductors}.
\newblock \emph{\bibinfo{journal}{J. Phys. Chem. Lett.}}
  \bibinfo{pages}{1421--1426} (\bibinfo{year}{2014}).

\bibitem{Mosconi:2013}
\bibinfo{author}{Mosconi, E.}, \bibinfo{author}{Amat, A.},
  \bibinfo{author}{Nazeeruddin, M.~K.}, \bibinfo{author}{Gr{\"a}tzel, M.} \&
  \bibinfo{author}{De~Angelis, F.}
\newblock \bibinfo{title}{First-principles modeling of mixed halide organometal
  perovskites for photovoltaic applications}.
\newblock \emph{\bibinfo{journal}{J. Phys. Chem. C}}
  \textbf{\bibinfo{volume}{117}}, \bibinfo{pages}{13902--13913}
  (\bibinfo{year}{2013}).

\bibitem{Liu:2013}
\bibinfo{author}{Liu, M.}, \bibinfo{author}{Johnston, M.~B.} \&
  \bibinfo{author}{Snaith, H.~J.}
\newblock \bibinfo{title}{Efficient planar heterojunction perovskite solar
  cells by vapour deposition}.
\newblock \emph{\bibinfo{journal}{Nature}} \textbf{\bibinfo{volume}{501}},
  \bibinfo{pages}{395--398} (\bibinfo{year}{2013}).

\bibitem{Xing:2013}
\bibinfo{author}{Xing, G.} \emph{et~al.}
\newblock \bibinfo{title}{Long-range balanced electron- and hole-transport
  lengths in organic-inorganic CH$_{3}$NH$_{3}$PbI$_{3}$}.
\newblock \emph{\bibinfo{journal}{Science}} \textbf{\bibinfo{volume}{342}}
  (\bibinfo{year}{2013}).

\bibitem{Yin:2014}
\bibinfo{author}{Yin, W.-J.}, \bibinfo{author}{Shi, T.} \&
  \bibinfo{author}{Yan, Y.}
\newblock \bibinfo{title}{Unusual defect physics in CH$_{3}$NH$_{3}$PbI$_{3}$ perovskite solar
  cell absorber}.
\newblock \emph{\bibinfo{journal}{Appl. Phys. Lett.}}
  \textbf{\bibinfo{volume}{104}}, \bibinfo{pages}{--} (\bibinfo{year}{2014}).

\bibitem{Kim:2014}
\bibinfo{author}{Kim, J.}, \bibinfo{author}{Lee, S.-H.}, \bibinfo{author}{Lee,
  J.~H.} \& \bibinfo{author}{Hong, K.-H.}
\newblock \bibinfo{title}{The role of intrinsic defects in methylammonium lead
  iodide perovskite}.
\newblock \emph{\bibinfo{journal}{J. Phys. Chem. Lett.}}
  \bibinfo{pages}{1312--1317} (\bibinfo{year}{2014}).

\bibitem{Brivio:2014}
\bibinfo{author}{Brivio, F.}, \bibinfo{author}{Butler, K.~T.},
  \bibinfo{author}{Walsh, A.} \& \bibinfo{author}{van Schilfgaarde, M.}
\newblock \bibinfo{title}{Relativistic quasiparticle self-consistent electronic
  structure of hybrid halide perovskite photovoltaic absorbers}.
\newblock \emph{\bibinfo{journal}{Phys. Rev. B}} \textbf{\bibinfo{volume}{89}},
  \bibinfo{pages}{155204} (\bibinfo{year}{2014}).

\bibitem{Umari:2014}
\bibinfo{author}{Umari, P.}, \bibinfo{author}{Mosconi, E.} \&
  \bibinfo{author}{De~Angelis, F.}
\newblock \bibinfo{title}{Relativistic gw calculations on CH$_{3}$NH$_{3}$PbI$_{3}$ and
  CH$_{3}$NH$_{3}$SnI$_{3}$ perovskites for solar cell applications}.
\newblock \emph{\bibinfo{journal}{Sci. Rep.}} \textbf{\bibinfo{volume}{4}}
  (\bibinfo{year}{2014}).

\bibitem{Baikie:2013}
\bibinfo{author}{Baikie, T.} \emph{et~al.}
\newblock \bibinfo{title}{Synthesis and crystal chemistry of the hybrid
  perovskite (CH3NH3)PbI3 for solid-state sensitised solar cell applications}.
\newblock \emph{\bibinfo{journal}{J. Mater. Chem. A}}
  \textbf{\bibinfo{volume}{1}}, \bibinfo{pages}{5628--5641}
  (\bibinfo{year}{2013}).

\bibitem{Knob:1990}
\bibinfo{author}{Knop, O.}, \bibinfo{author}{Wasylishen, R.~E.},
  \bibinfo{author}{White, M.~A.}, \bibinfo{author}{Cameron, T.~S.} \&
  \bibinfo{author}{Oort, M. J. M.~V.}
\newblock \bibinfo{title}{Alkylammonium lead halides. part 2. CH3NH3PbX3 (X=Cl,
  Br, I) perovskites: cuboctahedral halide cages with isotropic cation
  reorientation}.
\newblock \emph{\bibinfo{journal}{Can. J. Chem.}}
  \textbf{\bibinfo{volume}{68}}, \bibinfo{pages}{412--422}
  (\bibinfo{year}{1990}).

\bibitem{Yun:2014}
\bibinfo{author}{Wang, Y.} \emph{et~al.}
\newblock \bibinfo{title}{Density functional theory analysis of structural and
  electronic properties of orthorhombic perovskite CH$_{3}$NH$_{3}$PbI$_{3}$}.
\newblock \emph{\bibinfo{journal}{Phys. Chem. Chem. Phys.}}
  \textbf{\bibinfo{volume}{16}}, \bibinfo{pages}{1424--1429}
  (\bibinfo{year}{2014}).

\bibitem{PhysRevB.90.045207}
\bibinfo{author}{Men\'endez-Proupin, E.}, \bibinfo{author}{Palacios, P.},
  \bibinfo{author}{Wahn\'on, P.} \& \bibinfo{author}{Conesa, J.~C.}
\newblock \bibinfo{title}{Self-consistent relativistic band structure of the
  $\mathrm{CH}{}_{3}\mathrm{NH}{}_{3}\mathrm{PbI}{}_{3}$ perovskite}.
\newblock \emph{\bibinfo{journal}{Phys. Rev. B}} \textbf{\bibinfo{volume}{90}},
  \bibinfo{pages}{045207} (\bibinfo{year}{2014}).
\newblock \urlprefix\url{http://link.aps.org/doi/10.1103/PhysRevB.90.045207}.

\bibitem{Egger:2014}
\bibinfo{author}{Egger, D.~A.} \& \bibinfo{author}{Kronik, L.}
\newblock \bibinfo{title}{Role of dispersive interactions in determining
  structural properties of organic$-$inorganic halide perovskites: Insights
  from first-principles calculations}.
\newblock \emph{\bibinfo{journal}{J. Phys. Chem. Lett.}}
  \textbf{\bibinfo{volume}{5}}, \bibinfo{pages}{2728--2733}
  (\bibinfo{year}{2014}).

\bibitem{Kronik:2014}
\bibinfo{author}{Kronik, L.} \& \bibinfo{author}{Tkatchenko, A.}
\newblock \bibinfo{title}{Understanding molecular crystals with
  dispersion-inclusive density functional theory: Pairwise corrections and
  beyond}.
\newblock \emph{\bibinfo{journal}{Acc. Chem. Res.}}  (\bibinfo{year}{2014}).

\bibitem{Even:2013}
\bibinfo{author}{Even, J.}, \bibinfo{author}{Pedesseau, L.},
  \bibinfo{author}{Jancu, J.-M.} \& \bibinfo{author}{Katan, C.}
\newblock \bibinfo{title}{Importance of spin–orbit coupling in hybrid
  organic/inorganic perovskites for photovoltaic applications}.
\newblock \emph{\bibinfo{journal}{J. Phys. Chem. Lett.}}
  \textbf{\bibinfo{volume}{4}}, \bibinfo{pages}{2999--3005}
  (\bibinfo{year}{2013}).

\bibitem{Even:2014}
\bibinfo{author}{Even, J.}, \bibinfo{author}{Pedesseau, L.},
  \bibinfo{author}{Jancu, J.-M.} \& \bibinfo{author}{Katan, C.}
\newblock \bibinfo{title}{DFT and KP modelling of the phase transitions of lead
  and tin halide perovskites for photovoltaic cells}.
\newblock \emph{\bibinfo{journal}{physica status solidi (RRL) – Rapid
  Research Letters}} \textbf{\bibinfo{volume}{8}}, \bibinfo{pages}{31--35}
  (\bibinfo{year}{2014}).

\bibitem{Feng:2014}
\bibinfo{author}{Feng, J.} \& \bibinfo{author}{Xiao, B.}
\newblock \bibinfo{title}{Crystal structures, optical properties, and effective
  mass tensors of CH$_{3}$NH$_{3}$PbX$_{3}$ (X = I and Br) phases predicted from HSE06}.
\newblock \emph{\bibinfo{journal}{J. Phys. Chem. Lett.}}
  \textbf{\bibinfo{volume}{5}}, \bibinfo{pages}{1278--1282}
  (\bibinfo{year}{2014}).

\bibitem{Giorgi:2014}
\bibinfo{author}{Giorgi, G.}, \bibinfo{author}{Fujisawa, J.-I.},
  \bibinfo{author}{Segawa, H.} \& \bibinfo{author}{Yamashita, K.}
\newblock \bibinfo{title}{Cation role in structural and electronic properties
  of 3d organic-inorganic halide perovskites: A DFT analysis}.
\newblock \emph{\bibinfo{journal}{J. Phys. Chem. C}}
  \textbf{\bibinfo{volume}{118}}, \bibinfo{pages}{12176--12183}
  (\bibinfo{year}{2014}).

\bibitem{Lee_Teuscher_Miyasaka_Murakami_Snaith_2012}
\bibinfo{author}{Lee, M.~M.}, \bibinfo{author}{Teuscher, J.},
  \bibinfo{author}{Miyasaka, T.}, \bibinfo{author}{Murakami, T.~N.} \&
  \bibinfo{author}{Snaith, H.~J.}
\newblock \bibinfo{title}{Efficient hybrid solar cells based on
  meso-superstructured organometal halide perovskites}.
\newblock \emph{\bibinfo{journal}{Science}} \textbf{\bibinfo{volume}{338}},
  \bibinfo{pages}{643--647} (\bibinfo{year}{2012}).

\bibitem{Perd96}
\bibinfo{author}{Perdew, J.~P.}, \bibinfo{author}{Burke, K.} \&
  \bibinfo{author}{Ernzerhof, M.}
\newblock \bibinfo{title}{Generalized gradient approximation made simple}.
\newblock \emph{\bibinfo{journal}{Phys. Rev. Lett.}}
  \textbf{\bibinfo{volume}{77}}, \bibinfo{pages}{3865--3868}
  (\bibinfo{year}{1996}).

\bibitem{PhysRevLett.102.073005}
\bibinfo{author}{Tkatchenko, A.} \& \bibinfo{author}{Scheffler, M.}
\newblock \bibinfo{title}{Accurate molecular van der waals interactions from
  ground-state electron density and free-atom reference data}.
\newblock \emph{\bibinfo{journal}{Phys. Rev. Lett.}}
  \textbf{\bibinfo{volume}{102}}, \bibinfo{pages}{073005}
  (\bibinfo{year}{2009}).

\bibitem{Poglitsch1987}
\bibinfo{author}{Poglitsch, A.} \& \bibinfo{author}{Weber, D.}
\newblock \bibinfo{title}{Dynamic disorder in
  methylammoniumtrihalogenoplumbates (ii) observed by millimeter-wave
  spectroscopy}.
\newblock \emph{\bibinfo{journal}{J. Chem. Phys.}}
  \textbf{\bibinfo{volume}{87}}, \bibinfo{pages}{6373} (\bibinfo{year}{1987}).

\bibitem{Brivio:2013}
\bibinfo{author}{Brivio, F.}, \bibinfo{author}{Walker, A.~B.} \&
  \bibinfo{author}{Walsh, A.}
\newblock \bibinfo{title}{Structural and electronic properties of hybrid
  perovskites for high-efficiency thin-film photovoltaics from
  first-principles}.
\newblock \emph{\bibinfo{journal}{APL Materials}} \textbf{\bibinfo{volume}{1}},
  \bibinfo{pages}{--} (\bibinfo{year}{2013}).

\bibitem{Stoumpos:2013}
\bibinfo{author}{Stoumpos, C.~C.}, \bibinfo{author}{Malliakas, C.~D.} \&
  \bibinfo{author}{Kanatzidis, M.~G.}
\newblock \bibinfo{title}{Semiconducting tin and lead iodide perovskites with
  organic cations: phase transitions, high mobilities, and near-infrared
  photoluminescent properties}.
\newblock \emph{\bibinfo{journal}{Inorg. Chem.}} \textbf{\bibinfo{volume}{52}},
  \bibinfo{pages}{9019--9038} (\bibinfo{year}{2013}).

\bibitem{Kojima:2009}
\bibinfo{author}{Kojima, A.}, \bibinfo{author}{Teshima, K.},
  \bibinfo{author}{Shirai, Y.} \& \bibinfo{author}{Miyasaka, T.}
\newblock \bibinfo{title}{Organometal halide perovskites as visible-light
  sensitizers for photovoltaic cells}.
\newblock \emph{\bibinfo{journal}{J. Am. Chem. Soc.}}
  \textbf{\bibinfo{volume}{131}}, \bibinfo{pages}{6050--6051}
  (\bibinfo{year}{2009}).

\bibitem{El-Mellouhi:2013}
\bibinfo{author}{El-Mellouhi, F.}, \bibinfo{author}{Brothers, E.~N.},
  \bibinfo{author}{Lucero, M.~J.}, \bibinfo{author}{Bulik, I.~W.} \&
  \bibinfo{author}{Scuseria, G.~E.}
\newblock \bibinfo{title}{Structural phase transitions of the metal oxide
  perovskites SrTiO3, LaAlO3, and LaTiO3 studied with a screened hybrid
  functional}.
\newblock \emph{\bibinfo{journal}{Phys. Rev. B}} \textbf{\bibinfo{volume}{87}},
  \bibinfo{pages}{035107} (\bibinfo{year}{2013}).

\bibitem{Horzum:2013}
\bibinfo{author}{Horzum, S.} \emph{et~al.}
\newblock \bibinfo{title}{Phonon softening and direct to indirect band gap
  crossover in strained single-layer MoSe2}.
\newblock \emph{\bibinfo{journal}{Phys. Rev. B}} \textbf{\bibinfo{volume}{87}},
  \bibinfo{pages}{125415} (\bibinfo{year}{2013}).

\bibitem{JCCJCC20495}
\bibinfo{author}{Grimme, S.}
\newblock \bibinfo{title}{Semiempirical GGA-type density functional constructed
  with a long-range dispersion correction}.
\newblock \emph{\bibinfo{journal}{J. Comput. Chem.}}
  \textbf{\bibinfo{volume}{27}}, \bibinfo{pages}{1787--1799}
  (\bibinfo{year}{2006}).

\bibitem{PhysRevLett.108.236402}
\bibinfo{author}{Tkatchenko, A.}, \bibinfo{author}{DiStasio, R.~A.},
  \bibinfo{author}{Car, R.} \& \bibinfo{author}{Scheffler, M.}
\newblock \bibinfo{title}{Accurate and efficient method for many-body van der
  Waals interactions}.
\newblock \emph{\bibinfo{journal}{Phys. Rev. Lett.}}
  \textbf{\bibinfo{volume}{108}}, \bibinfo{pages}{236402}
  (\bibinfo{year}{2012}).

\bibitem{Krukau:2006}
\bibinfo{author}{Krukau, A.~V.}, \bibinfo{author}{Vydrov, O.~A.},
  \bibinfo{author}{Izmaylov, A.~F.} \& \bibinfo{author}{Scuseria, G.~E.}
\newblock \bibinfo{title}{Influence of the exchange screening parameter on the
  performance of screened hybrid functionals}.
\newblock \emph{\bibinfo{journal}{J. Chem. Phys.}}
  \textbf{\bibinfo{volume}{125}} (\bibinfo{year}{2006}).

\bibitem{Frost:2014}
\bibinfo{author}{Frost, J.~M.} \emph{et~al.}
\newblock \bibinfo{title}{Atomistic origins of high-performance in hybrid
  halide perovskite solar cells}.
\newblock \emph{\bibinfo{journal}{Nano Lett.}} \textbf{\bibinfo{volume}{14}},
  \bibinfo{pages}{2584--2590} (\bibinfo{year}{2014}).

\bibitem{Mosconi:2014}
\bibinfo{author}{Mosconi, E.}, \bibinfo{author}{Quarti, C.},
  \bibinfo{author}{Ivanovska, T.}, \bibinfo{author}{Ruani, G.} \&
  \bibinfo{author}{De~Angelis, F.}
\newblock \bibinfo{title}{Structural and electronic properties of organo-halide
  lead perovskites: a combined IR-spectroscopy and ab initio molecular dynamics
  investigation}.
\newblock \emph{\bibinfo{journal}{Phys. Chem. Chem. Phys.}}
  (\bibinfo{year}{2014}).

\bibitem{Even:2014b}
\bibinfo{author}{Even, J.}, \bibinfo{author}{Pedesseau, L.} \&
  \bibinfo{author}{Katan, C.}
\newblock \bibinfo{title}{Analysis of multivalley and multibandgap absorption
  and enhancement of free carriers related to exciton screening in hybrid
  perovskites}.
\newblock \emph{\bibinfo{journal}{J. Phys. Chem. C}}
  \textbf{\bibinfo{volume}{118}}, \bibinfo{pages}{11566--11572}
  (\bibinfo{year}{2014}).

\bibitem{Burschka:2013}
\bibinfo{author}{Burschka, J.} \emph{et~al.}
\newblock \bibinfo{title}{Sequential deposition as a route to high-performance
  perovskite-sensitized solar cells}.
\newblock \emph{\bibinfo{journal}{Nature}} \textbf{\bibinfo{volume}{499}},
  \bibinfo{pages}{316--319} (\bibinfo{year}{2013}).

\bibitem{C3TA13606J}
\bibinfo{author}{Niu, G.} \emph{et~al.}
\newblock \bibinfo{title}{Study on the stability of CH$_{3}$NH$_{3}$PbI$_{3}$ films and the
  effect of post-modification by aluminum oxide in all-solid-state hybrid solar
  cells}.
\newblock \emph{\bibinfo{journal}{J. Mater. Chem. A}}
  \textbf{\bibinfo{volume}{2}}, \bibinfo{pages}{705--710}
  (\bibinfo{year}{2014}).

\bibitem{doi:10.1021.jz500434p}
\bibinfo{author}{Wehrenfennig, C.}, \bibinfo{author}{Liu, M.},
  \bibinfo{author}{Snaith, H.~J.}, \bibinfo{author}{Johnston, M.~B.} \&
  \bibinfo{author}{Herz, L.~M.}
\newblock \bibinfo{title}{Homogeneous emission line broadening in the organo
  lead halide perovskite CH3NH3PbI3–XClX}.
\newblock \emph{\bibinfo{journal}{J. Phys. Chem. Lett.}}
  \textbf{\bibinfo{volume}{5}}, \bibinfo{pages}{1300--1306}
  (\bibinfo{year}{2014}).

\bibitem{fahad_a}
\bibinfo{author}{Loferski, J.~J.}
\newblock \bibinfo{title}{Theoretical considerations governing the choice of
  the optimum semiconductor for photovoltaic solar energy conversion}.
\newblock \emph{\bibinfo{journal}{Journal of Applied Physics}}
  \textbf{\bibinfo{volume}{27}}, \bibinfo{pages}{777--784}
  (\bibinfo{year}{1956}).

\bibitem{fahad_b}
\bibinfo{author}{Alharbi, F.} \emph{et~al.}
\newblock \bibinfo{title}{Abundant non-toxic materials for thin film solar
  cells: Alternative to conventional materials}.
\newblock \emph{\bibinfo{journal}{Renew. Energ.}}
  \textbf{\bibinfo{volume}{36}}, \bibinfo{pages}{2753 -- 2758}
  (\bibinfo{year}{2011}).

\bibitem{fahad_c}
\bibinfo{author}{Zakutayev, A.} \emph{et~al.}
\newblock \bibinfo{title}{Defect tolerant semiconductors for solar energy
  conversion}.
\newblock \emph{\bibinfo{journal}{J. Phys. Chem. Lett.}}
  \textbf{\bibinfo{volume}{5}}, \bibinfo{pages}{1117--1125}
  (\bibinfo{year}{2014}).

\bibitem{fahad_d}
\bibinfo{author}{Fonash, S.}
\newblock \emph{\bibinfo{title}{Solar Cell Device Physics}}
  (\bibinfo{publisher}{Oxford: Academic Press}, \bibinfo{year}{2010}).

\bibitem{fahad_e}
\bibinfo{author}{Law, M.}, \bibinfo{author}{Solley, E.},
  \bibinfo{author}{Liang, M.} \& \bibinfo{author}{Burk, D.}
\newblock \bibinfo{title}{Self-consistent model of minority-carrier lifetime,
  diffusion length, and mobility}.
\newblock \emph{\bibinfo{journal}{Electron Device Lett.}}
  \textbf{\bibinfo{volume}{12}}, \bibinfo{pages}{401--403}
  (\bibinfo{year}{1991}).

\bibitem{fahad_f}
\bibinfo{author}{Gr{\"a}tzel, M.}
\newblock \bibinfo{title}{Photoelectrochemical cells}.
\newblock \emph{\bibinfo{journal}{Nature}} \textbf{\bibinfo{volume}{414}},
  \bibinfo{pages}{338--344} (\bibinfo{year}{2001}).

\bibitem{Choi:2013}
\bibinfo{author}{Choi, J.~J.}, \bibinfo{author}{Yang, X.},
  \bibinfo{author}{Norman, Z.~M.}, \bibinfo{author}{Billinge, S. J.~L.} \&
  \bibinfo{author}{Owen, J.~S.}
\newblock \bibinfo{title}{Structure of methylammonium lead iodide within
  mesoporous titanium dioxide: Active material in high-performance perovskite
  solar cells}.
\newblock \emph{\bibinfo{journal}{Nano Lett.}} \textbf{\bibinfo{volume}{14}},
  \bibinfo{pages}{127--133} (\bibinfo{year}{2013}).

\bibitem{Stranks:2013}
\bibinfo{author}{Stranks, S.~D.} \emph{et~al.}
\newblock \bibinfo{title}{Electron-hole diffusion lengths exceeding 1
  micrometer in an organometal trihalide perovskite absorber}.
\newblock \emph{\bibinfo{journal}{Science}} \textbf{\bibinfo{volume}{342}},
  \bibinfo{pages}{341--344} (\bibinfo{year}{2013}).

\bibitem{PhysRevB.13.5188}
\bibinfo{author}{Monkhorst, H.~J.} \& \bibinfo{author}{Pack, J.~D.}
\newblock \bibinfo{title}{Special points for brillouin-zone integrations}.
\newblock \emph{\bibinfo{journal}{Phys. Rev. B}} \textbf{\bibinfo{volume}{13}},
  \bibinfo{pages}{5188--5192} (\bibinfo{year}{1976}).

\bibitem{numrec97}
\bibinfo{author}{Press, W.~H.}, \bibinfo{author}{Teukolsky, S.~A.},
  \bibinfo{author}{Vetterling, W.~T.} \& \bibinfo{author}{Flannery, B.~P.}
\newblock \emph{\bibinfo{title}{Numerical recipes}}
  (\bibinfo{publisher}{Cambridge University Press}, \bibinfo{year}{1997}),
  \bibinfo{edition}{3} edn.

\bibitem{g09}
\bibinfo{author}{Frisch, M.~J.} \emph{et~al.}
\newblock \bibinfo{title}{Gaussian~09 revision \{B\}.01}.
\newblock \bibinfo{note}{Gaussian Inc. Wallingford CT 2009}.

\bibitem{Kudin:2000}
\bibinfo{author}{Kudin, K.~N.} \& \bibinfo{author}{Scuseria, G.~E.}
\newblock \bibinfo{title}{Linear-scaling density-functional theory with
  gaussian orbitals and periodic boundary conditions: Efficient evaluation of
  energy and forces via the fast multipole method}.
\newblock \emph{\bibinfo{journal}{Phys. Rev. B}} \textbf{\bibinfo{volume}{61}},
  \bibinfo{pages}{16440--16453} (\bibinfo{year}{2000}).

\bibitem{Weigend:2005}
\bibinfo{author}{Weigend, F.} \& \bibinfo{author}{Ahlrichs, R.}
\newblock \bibinfo{title}{Balanced basis sets of split valence, triple zeta
  valence and quadruple zeta valence quality for H to Rn: Design and assessment
  of accuracy}.
\newblock \emph{\bibinfo{journal}{Phys. Chem. Chem. Phys.}}
  \textbf{\bibinfo{volume}{7}}, \bibinfo{pages}{3297--3305}
  (\bibinfo{year}{2005}).

\bibitem{El-Mellouhi:2011}
\bibinfo{author}{El-Mellouhi, F.}, \bibinfo{author}{Brothers, E.~N.},
  \bibinfo{author}{Lucero, M.~J.} \& \bibinfo{author}{Scuseria, G.~E.}
\newblock \bibinfo{title}{Modeling of the cubic and antiferrodistortive phases
  of SrTiO$_{3}$ with screened hybrid density functional theory}.
\newblock \emph{\bibinfo{journal}{Phys. Rev. B}} \textbf{\bibinfo{volume}{84}},
  \bibinfo{pages}{115122} (\bibinfo{year}{2011}).

\bibitem{phonopy}
\bibinfo{author}{Togo, A.}, \bibinfo{author}{Oba, F.} \&
  \bibinfo{author}{Tanaka, I.}
\newblock \bibinfo{title}{First-principles calculations of the ferroelastic
  transition between rutile-type and ${\text{CaCl}}_{2}$-type
  ${\text{SiO}}_{2}$ at high pressures}.
\newblock \emph{\bibinfo{journal}{Phys. Rev. B}} \textbf{\bibinfo{volume}{78}},
  \bibinfo{pages}{134106} (\bibinfo{year}{2008}).

\end{thebibliography}
\end{document}